\documentclass[final,5p,times,twocolumn]{elsarticle}
\usepackage{graphicx}
\usepackage{amsmath,amsfonts,amssymb,bm,ulem}
\usepackage[usenames]{color}

\newcommand{\g}{eps}
\journal{Solid State Communications}
 \begin{document} 
\begin{frontmatter}
\title{Slow relaxations of magnetoresistance in AlGaAs-GaAs quantum
  well structures quenched in a magnetic field}
\author[a]{N. V.~Agrinskaya}
\author[a]{V. I. Kozub}
\author[a]{D. V. Shamshur}
\author[a]{A. V.~Shumilin}
\author[a,b]{Y. M. Galperin}

\address[a]{A. F. Ioffe Institute, Russian Academy of Sciences,
  194021 Saint Petersburg, Russia}
\address[b]{Department of Physics, University of Oslo, PO Box 1048
  Blindern, 0316 Oslo, Norway}

\begin{abstract}

We observed a slow relaxation of magnetoresistance in response to
applied magnetic field in selectively doped p-GaAs-AlGaAs structures
with partially filled upper Hubbard band.  We have paid a special
attention to exclude the effects related to temperature fluctuations.
Though this  effect is important, we have found that the general
features of slow relaxation still persist. This  behavior is
interpreted as related to the properties of the Coulomb glass formed
by charged centers with account of spin correlations, which are
sensitive to an external magnetic field.  Variation of the magnetic
field changes numbers of impurity complexes of different types. As a
result, it effects the shape and depth  of the polaron gap formed at
the states belonging to the percolation cluster responsible for the
conductance. The suggested model explains both the qualitative
behavior and the order of magnitude of the slowly relaxing
magnetoresistance. \\[0.1in]

\end{abstract}


\end{frontmatter}

\section{Introduction}

Recently~\cite{ours} we reported observation of long time
relaxation of (negative) magnetoresistance in p-type AlGaAs-GaAs
quantum well structures where both wells and barriers were doped
by Be. We argued that in these structures so-called $A^+$ centers
-- doubly occupied acceptors belonging to the upper Hubbard band
-- are formed in the well. The observed long time behavior of
magnetoresistance was explained by  polaron effects involving
the spin-correlated $A^+$-centers~\cite{end1}.

In the present paper, we report more detailed studies of slow
relaxations in the same samples. The main point requiring more
detailed measurements is that in our previous paper~\cite{ours}
fluctuations of the sample temperature during measurements were out of
proper control and account. Since temperature fluctuations may lead to
a variation of the sample resistance our previous results and their
interpretation need a proper verification.  We are grateful to
Z.~Ovadyahy who attracted our attention to this problem.

Investigation of role of the temperature fluctuations is the main goal
of the present work. We will show that, although the temperature
fluctuations indeed can produce a pronounced effect on the resistance,
the observed long time relaxations of magnetoresistance have an
independent source similar to that mechanism considered
in~\cite{ours}. More careful analysis of the experimental results has
required, however, some modification of the theoretical
model~\cite{ours}.  Namely, now we believe that two mechanisms can
lead to a slow relaxation of the magnetoresistance. The fist one is
the magnetic-field-induced shift of the chemical potential of the
holes~\cite{ours}.  The second mechanism is due to a direct influence
of the magnetic field on the \textit{aggregates} responsible for
formation of the polaron gap at the sites responsible for the
conductance~\cite{ours1}. Contrary to~\cite{ours1}, we conclude that
it is the second mechanism that explains the refined experimental
results.

The paper is organized as follows. In Sec.~\ref{samples} we briefly
describe the samples and experimental procedure and report the
experimental results. These results are interpreted in
Sec.~\ref{discussion} where the theoretical models are considered and
compared with experiment.

\section{Samples, experimental procedure, and results}\label{samples}

We used the structures containing 10 GaAs quantum wells separated by
Al$_{0.3}$Ga$_{0.7}$As barriers. Thickness of both wells and barriers
was 15 nm. Confining Al$_{0.3}$Ga$_{0.7}$As layers had thickness of 20
nm. The growth procedure is described in detail
in~\cite{ours}. Central regions with thickness of 5 nm of both wells
and barriers were $p$-doped with Be (concentration $10^{17}$
atoms/cm$^3$). The contacts were made by 2 min burning at 450$^\circ$
C in deposited gold containing 3\% of Zn. The samples were shaped as
Hall bars. The resistance was determined from the voltage between the
voltage probes at a fixed current of 1 nA.  
We studied several samples cut from the same
wafer. The samples were relatively low-Ohmic ($10^5 -
10^6$~Ohms/$\square$ at 4~K) that is, in our opinion, due to the fact
that the impurity band formed by $A^+$ centers is rather close to the
valence band, the binding energy being $ \sim 10$ meV~\cite{ours}.  We
have checked that for all our measurements the $I-V$ curves were
linear in the temperature domain 4.2-1.35 K. The temperature
dependence of the resistance is compatible with the 2D Mott variable
range hopping (see Fig.~1 in~\cite{ours}), $R \propto \exp
(T_0/T)^{1/3}$, with $T_0=1000$ K for the sample under
investigation. The magnetoresistance curves were close to those
reported in~\cite{ours}, namely, the linear negative magnetoresistance
crossed over to the quadratic positive magnetoresistance at stronger
magnetic fields.

Studies of the slow relaxations induced by controlled variation of
external magnetic field are specifically difficult because our
thermometers are sensitive to magnetic field. Therefore, to investigate
role of the temperature fluctuations we made special measurements of the
sample resistance and its temperature in the absence of magnetic field.
\begin{figure}[h]
\centerline{ \includegraphics[width=8cm]{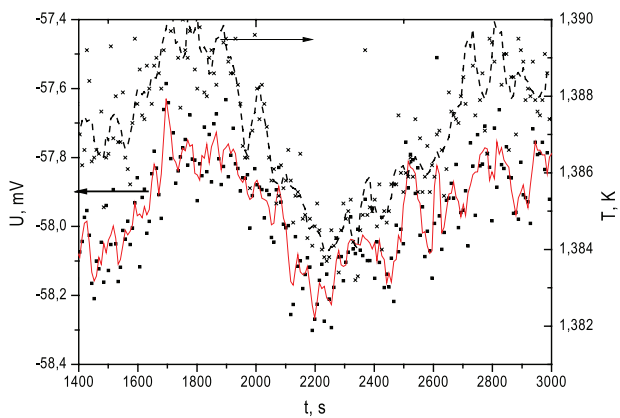} }
\caption{Evolution of the voltage across the sample and its
temperature with time at $H=0$ The dashed curve corresponds to
evolution of temperature while the solid curve - to evolution of
voltage. The curves were drawn by the adjacent averaging of
the experimental points (involving 10 points).  \label{fig1} }
\end{figure}
 The corresponding curves are presented in~Fig.~1. One
concludes that temperature fluctuations up to several mK indeed exist.
We believe that these fluctuations are due to fluctuations of the
pressure in the system pumped by a pre-evacuation pump without a
sufficiently large damping volume, which should be at least by an order
of magnitude larger than the Dewar volume. It is also seen that the
resistance fluctuates keeping time with the temperature fluctuations --
there is no lag between variations of the resistance and temperature.
This fact allows us to conclude that at a given time the resistance of
the sample is controlled by the temperature of the sample measured at
the same instant of time. Accordingly, we can extract temperature
derivative of the resistance and then estimate the corresponding
fluctuation-induced contribution to the zero-field resistance at a known
temperature as
\begin{equation}\label{correction} \Delta R \simeq
\frac{\partial R}{\partial T}\Delta T\, . \end{equation}
{}From Fig.~1
it is seen that the variation of temperature by 1~mK leads to a
variation of voltage and thus the variation of resistance is of the
order of 0.2\%.

\begin{figure}[h] \centerline{ \includegraphics[width=8cm]{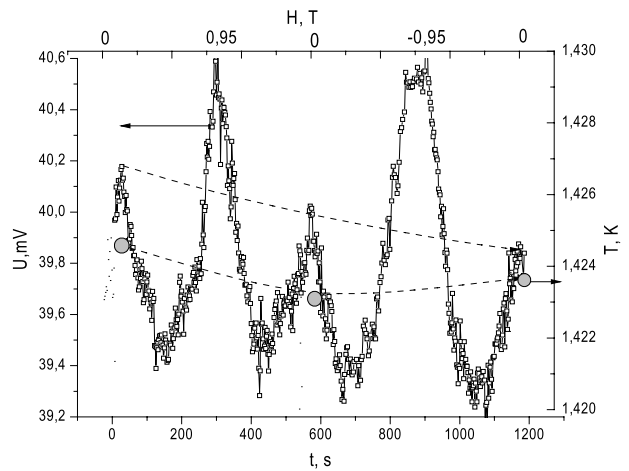} }
\caption{Magnetoresistance as a function of time. Small open
squares represent the voltage across the sample while the circles
represent the temperatures at the instants of time when $H = 0$.
Dashed lines connect the values of voltage and temperature at the
times when $H=0$. \label{fig2} }
 \end{figure}

Shown in Fig.~2 is time dependence of magnetoresistance (or rather
of the voltage between the probe contacts) obtained by ramping of
magnetic field in time with very slow sweep rate (about 30 Oe/s ).
Large circles show measured temperatures at $H=0$, that correspond
to 3 instants of time in Fig.~\ref{fig2}. Note that the
magnetoresistance curves of Fig.~\ref{fig2} were close to those
reported in Fig.~1 of our previous paper,~\cite{ours} which were
obtained at fast magnetic field sweep of $ \sim 200$ Oe/s
\cite{end2}. 
 Namely, the
linear negative magnetoresistance crossed over to the quadratic
positive magnetoresistance at stronger magnetic fields. 
As it is seen from
Fig.~\ref{fig2}, the resistance at the successive instants of time when
$H = 0$ is different -- it gradually decreases with time (by $\approx
1$\%). At the same time, the temperature at that instants is almost the
same showing initially weak decrease and then a weak increase within the
interval not exceeding $\approx 1$~mK. These deviations of temperature
would lead only to deviations of the resistance by $\approx 0.2\%$
(upward and downward).
These measurements demonstrate that, in
addition to thermal effects, the resistance tends to decrease after
application of the magnetic field, the effect at the times $\sim 1000$~s
being at the level of 1\%.

In Fig.~3, we present the behavior of the voltage across the probe
contacts (and thus of the resistance) for two opposite current
directions for the sample quenched during 20 min in a steady magnetic
field of 0.7 T after subsequent switching off the field.
\begin{figure}[h] \centerline{
\includegraphics[width=8cm]{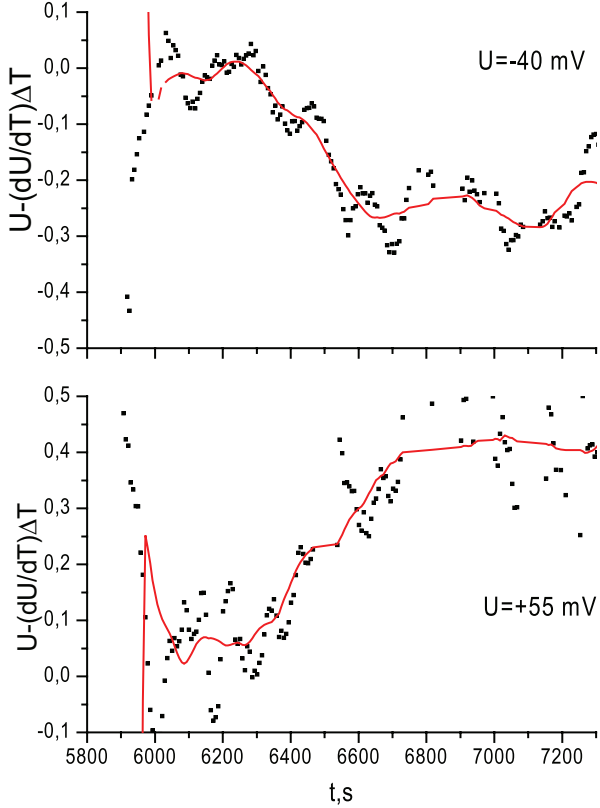} }
\caption{Evolution of sample resistance after switching off a
steady magnetic field. The points correspond to values of the
voltage for the two different current directions corrected for the
effect of temperature while the curves are obtained by averaging
procedure described above. The sharp peaks from the left
correspond to the instant when the magnetic field is switched off.}
\end{figure}
 In this case the temperature measurements
are available during all the time of the measurement. Thus we are able
to correct the resistance curve for the temperature fluctuations at all
experimental points. It is seen that the corrected voltage, $U -
(\partial U/\partial T) \Delta T$, slowly relaxes to higher values and
is saturated at times $\sim 800$~s. It is important that the saturation
value is equal to the resistance of the sample before application of the
magnetic field. In other words, after application of the magnetic field
the resistance slowly decreases with time but after switching off of the
field it is gradually restored to its initial value. It is also
important that the restoration time is nearly equal to the time at which
the sample was subjected to the external magnetic field.

\section{Discussion} \label{discussion}

We believe that the results
presented above evidence a specific slow response of the sample
resistance to applied magnetic field. This response cannot be explained
by temperature fluctuations, which would not lead to the observed
contribution to the resistance, which monotonously depends on
time.
 As it follows both from experiment and our estimates, a steady
heating due to eddy currents induced in the Cu substrate by variations
in the magnetic field is negligibly small. At the same time, the
observed resistance gradually decreases after application of magnetic
field and tends to restore after switching it off. In addition, as it
follows from direct measurements and our estimates, a possible effect of
temperature fluctuations would be about an order of magnitude less than
the observed variation of resistance.

\subsection{Theoretical model}

\paragraph*{Hole-impurity complexes. --}
Let us fist review the previously suggested mechanism of slow
relaxations~\cite{ours},  which is a generalization of the one developed
in~\cite{ours1}. We assume that some localized states
(including $A^+$-centers -- acceptor atoms doubly occupied by holes and
located in the wells) form bistable aggregates. The low energy states of these
 aggregates are almost degenerate, the transitions between the
states can take place only due to \textit{many-electron} processes that
results in very long transition times. Though the above aggregates do
not belong to the percolation cluster, they can be polarized by the
electrical charges located at the hopping sites. The \textit{polaron
clouds} formed by the aggregates for typical hopping sites lead to
formation of a \textit{polaron gap} at the Fermi level. As a result,
conductance decreases.

An important feature of the material under consideration is that, in
addition to doubly and singly occupied acceptors in the well ($A^+$
centers and $A^0$ centers, respectively), there can also exist
\textit{neutral} complexes consisting of negatively charged acceptor
in the barrier and a valence-band hole located in the well, which is
coupled to the acceptor due to Coulomb interaction.  We will call such
complexes the $\tilde A^0$ centers. Analog of such complexes in a
$n$-type conductor were introduced in~\cite{Larsen}, and in what
follows we will exploit the  scheme suggested in that paper.  In the
material under consideration, one has to consider acceptor complexes
of 4 types: (i) positively charged $A^+$-centers -- doubly occupied
(by holes) acceptors located in the wells and forming the upper
Hubbard band; (ii) neutral $A^0$-centers -- singly occupied (by holes)
acceptors located in the wells; (iii) negative $A^-$-centers --
acceptors in the barriers; and (iv) neutral $\tilde A^0$-complexes
consisting of an acceptor in the barrier coupled to a valence-band
hole in the well.

Assuming that he dopant concentrations, $N_A$, within the barrier and
within the well are equal one has
\begin{equation}\label{neut1} \underbrace{N_{A^+} +
N_{A^0}}_{\text{well}} =\underbrace{ N_{A^-} + N_{\tilde
A^0}}_{\text{barrier}}= N_A\, .
\end{equation} Due to charge conservation we also have
\begin{equation}\label{neut2} N_{A^+} = N_{A^-}\, .
\end{equation} In combination with Eq.~(\ref{neut1}) we get
\begin{equation}\label{neut3} N_{A^0} = N_{\tilde A^0}\, .
\end{equation}

Since there are three unknown concentrations, $N_i$, one needs one
more relation in order to calculate the chemical potential, $\mu$. To
derive this relation let us take into account the fact that a hole
provided by an acceptor within the barrier is inevitably captured
within the well.  However, it can belong either to an $A^+$ or it can
exist as a valence band hole coupled to a negative acceptor in the
barrier forming an $\tilde A^0$ center. In particular, a hole released
by an acceptor in the barrier can be captured by an occupied acceptor
in the well forming $A^+$ center provided
 \begin{equation} \label{A+} U_{A^+} + \frac{e^2}{\kappa \sqrt{r^2 +
(d_w+ d_b)^2/4}} \geq U_{\tilde A^0}\, .
\end{equation} Here $U_{i}$ is the binding energy of the center of
$i$th type, $r$ is a distance between the two acceptors along the
plane of the structure, $d_w$ and $d_b$ are the thicknesses of the
well and the barrier, respectively.
\begin{figure}[h] \centerline{
\includegraphics[width=4cm]{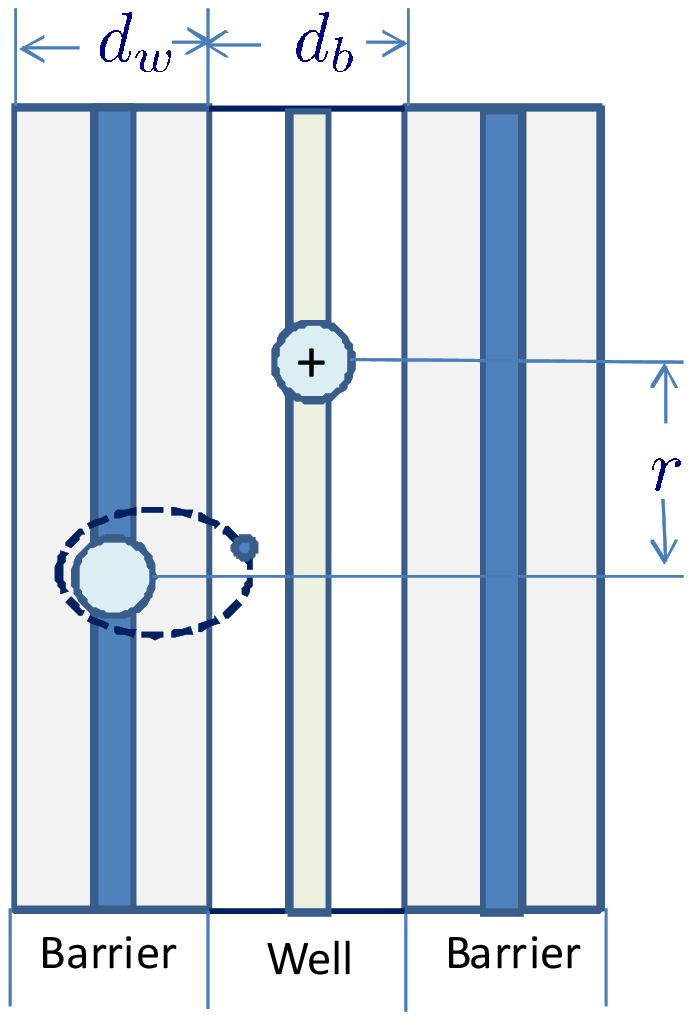}}
\caption{\label{fig4}}
\end{figure}
 For simplicity we assume that $kT \ll U_i$ and that the dopants form
$\delta$-layers in the middle of the well and the barrier. The
l.h.s. of Eq.~(\ref{A+}) gives the energy gain due to formation of an
$A^+$-center (the second term describes the interaction of the
$A^+$-center with the $A^-$-center within the barrier). One concludes
that if $U_{\tilde A^0} <U_{A^+}$ \textit{all} the acceptors within
the well become the $A^+$-centers (while the acceptors within the
barriers all become the $A^-$-centers).

In contrast, if \begin{equation}\label{A0} U_{\tilde A^0} - U_{A^+}
\geq\frac{2e^2}{\kappa (d_w + d_b)} \end{equation} both $A^+$-centers
and $\tilde A^-$-centers cannot be formed, and only $A^0$- and $\tilde
A^0$-centers are present. Strictly speaking, in both of these limits
the charge transport over the impurity band cannot take place. Such a
transport requires partial ``compensation" of the acceptors, which is
possible provided
     \begin{equation}\label{comp1} 0< U_{\tilde A^0} - U_{A^+} <
\frac{2e^2}{\kappa (d_w + d_b)}\, .
    \end{equation} In this case the relation between the
concentrations of the $A^+$- and $\tilde A^0$-centers (and thus of the
$A^0$-centers) depends on the dopant concentration, $N_A$, which
controls the typical value of $r$ in Eq.~(\ref{A+}). Note that if
Eq.~(\ref{A+}) holds for $r \simeq N_A^{-1/2}$ the most part of
acceptors within the well form $A^+$ centers while the number of $A^0$
centers is exponentially small (due to exponentially small probability
to find the values of $r > N_A^{-1/2}$).  The effective
``compensation" takes place if
    \begin{equation}\label{NA} 0< U_{\tilde A^0} - U_{A^+} <
\frac{e^2}{\kappa \sqrt{N_A^{-1} + (d_w + d_b)^2/4}}\,
. \end{equation} In this case the relation~(\ref{A+}) defines the
average distance between $A^+$ centers and thus the relative densities
of states of the upper Hubbard band (formed by  $A^+$- and
$A^0$-centers) and of the lower Hubbard band (formed by $\tilde A^0$-
and $A^-$-centers), $g_u$ and $g_l$, respectively. This distance,
$r_c$,corresponds to the equality relation in Eq.~(\ref{A+}).

\paragraph*{Influence of magnetic field. --}  The external magnetic
field tending to align  all the spins of the holes decreases the
binding energy of the $A^+$-centers, $U_{A^+}$, thus changing the
balance between $A^+$- and $\tilde A^0$-centers. As it is known for
$A^+$-centers in  GaAs/AlGaAs structures, the total spin of $A^+$
state is 2 while the spin of a separate hole is 3/2
(see~\cite{Averkiev}). Thus the application of the magnetic field $H$
leads to an effective increase of the Hubbard energy by the Zeeman
energy term $\mu_B g |H|$, $\mu_B$ being the Bohr magneton. As a
result, the holes redistribute between the acceptors and the chemical
potential is shifted as~\cite{Matveev}
\begin{equation} \delta \mu \sim \mu_B g |H |\frac{ g_u - g_l}{g_u +
g_l} \, .  \label{mu1}\end{equation}

\paragraph*{Estimate for magnetoresistance. -- }

To compare the experimental results with our theoretical model let us
reconstruct the estimate~\cite{ours} for magnetoresistance. Consider a
hopping site with equilibrium energy $\varepsilon$ (referred to the
Fermi level) coupled to some two-level system (TLS) with inter-level
splitting $E$ in the equilibrium. The coupling potential $U(R)$ (where
$R$ is a distance between the hopping site and the TLS) is defined as
the difference between the coupling energy for the two TLS states
corresponding to the upper and lower levels of the TLS. If one creates
an excitation at this hopping site (either an electron or a hole
depending on the sign of $\varepsilon$) the TLS splitting is changed
as $E \rightarrow E + U(R)$. If $U$ is negative and $|U| > E$, the TLS
changes its state with respect to the equilibrium
one. Correspondingly, the excitation energy is also changed as
$\varepsilon + U(R)$, and for an electron excitation its energy is
lowered. For the hole excitation the same effect will take place if $U
> 0$. As a result, the presence of a TLS leads to a formation of a
polaron gap around the Fermi level having width of $2U$ and depth
depending on the concentration of TLSs. The shape of the gap can be
found as follows~\cite{ours1}.  Specifying the density of states for
the TLSs with inter-level spacing $E$ and relaxation time $\tau$ as
$P(E, R, \tau)= \bar {P}/\tau$ where ${\bar P} = \mathrm{const}$ we
find the distribution function of polaron shifts, $U$, from the
equation

    \begin{eqnarray} {\cal F}(U)\, {\rm d} U &=& 2\pi R\,
    dR \, \bar{P} \int_{\tau_{\min}}^{\tau_{\max}} \frac{{\rm d} \tau}{
    \tau} \int_0^{U(R)} {\rm d} E \nonumber \\ &=&U(R) \bar{P}\, 2 \pi R\,
    dR \, \ln \frac{\tau_{\max}}{\tau_{\min}}\, .
    \end{eqnarray}
This is just the probability to find a TLS
 providing the polaron shift
between $U$ and $U+dU$ and located within the layer between $R$
and $R+dR$, which can switch between its states during the time of
experiment, $\tau_{\max}$.
Collecting contributions of all relevant
 TLSs, i.e., excluding all TLSs with polaron
shifts $\le U$, we obtain the shape of the polaron gap as
\begin{equation}\label{FU} F(U) = - 2 \pi \bar {P} \ln
\frac{\tau_{\max}}{\tau_{\min}} \int_{U}^{\infty} {\rm d} U'\, U' R(U')
\left(\frac{{\rm d} R}{{\rm d} U'}\right)\, . \end{equation} Following
our previous paper~\cite{ours1} let us consider many-electron
aggregates(electronic TLSs) 
 of chessboard type formed from the pairs with one
occupied and one empty site. These building blocks can be regarded as
pseudo-spins. The two states of the aggregate correspond then
corresponds to opposite direction of all of the ``spins". The transition
between the two states can be due to either a coherent multi-electron
hop, or due to motion of a ``domain wall" separating parts of the
aggregate with different phases of spin orientation. For this model, the size
of the optimal aggregate turns out to be relatively large and its
coupling to hopping sites is most effective when the hopping site is
close to one of the sites belonging to the aggregate. In this case the
coupling is just due to Coulomb interaction between the charges of the
hopping site and the nearest site of the aggregate. The coupling to the
rest sites of the aggregate turns out to be much weaker. In this case
the distribution function $\bar P$ should be multiplied by the number
$N$ of the sites belonging to the aggregate since the hopping site can
be coupled to any site within the aggregate. In what follows we will
absorb this factor into $\bar P$. For the Coulomb charge-charge
interaction, $U = e^2/\kappa R$, the integrand in Eq.~(\ref{FU}) is
$\propto U'^{-2}$ and, correspondingly, the integral is controlled by
its lower limit. The apparent divergence at $U \rightarrow 0$ has a
cut-off due to the fact that the small values of interaction energy $U$
correspond to large distances between the hopping site and the
aggregate. At large distance the Coulomb charge-charge coupling is
replaced by a much weaker dipole coupling originating from a (random)
dipole moment of the aggregate. Having this fact in mind we assume that
 the cut-off as the energy $\varepsilon_h$ corresponding to the typical
coupling energy between the sites separated by the typical hopping
length $r_h$. The quantity $\varepsilon_h$ is just the width of the
hopping energy band. Following this reasoning we assume that the gaps
with $U < \varepsilon_h$ do not effectively influence the hopping
transport. As a result, the density of states near the Fermi level as a
function of energy can be described by an interpolation relation as
    \begin{equation}\label{gap} \frac{\delta g_u}{g_u} \sim - \frac{2
    \pi\bar{P}}{\sqrt{\varepsilon^2 +
    \varepsilon_h^2}}\left(\frac{e^2}{\kappa}\right)^2 \ln
    \frac{\tau_{\max}}{\tau_{\min}} \, .
    \end{equation}
In the framework of
the model~\cite{ours1} the time scales $\tau_{\max/\min}$ are controlled
by the gate voltage protocol while In the present case case, the
external perturbation is due to variation of the magnetic field. An
instant application of magnetic field leads to a shift of the chemical
potential within the polaron gap.

Let us estimate the corresponding change in conductance assuming that
    $$\delta \mu \lesssim \mu_B g H \leq T \ll \varepsilon_h\, .$$
 {}From Eqs.~(\ref{gap}) and (\ref{mu1}) one can  expect that
energy shift of $\delta \varepsilon \sim \delta \mu \sim \mu_B g
|h|$ will lead to a relative shift in the density of states as
well as in the i conductance of the order of $(\delta \mu
/\varepsilon_h)^2=(\mu_B g H/\varepsilon_h))^2$. Therefore one can
expect that the magnetoresistance in an oscillating magnetic field
would  form a pattern following the magnetic field. In addition to
such a pattern, the experiments demonstrate a gradual increase of the
conductance with time.

To explain this increase we introduce an additional mechanism,
which is due to
a slow reconstruction of the polaron gap adjusting its
\textit{depth } to the magnetic-field-dependent number of the
``active" sites forming the polaron cloud. The reconstruction can
result from a \textit{direct} influence of the magnetic field on
the aggregates forming the polaron clouds. Indeed, the aggregates
are formed from pairs of sites allowing electron (hole)
transitions within the pairs. These transitions change
\textit{types} of the impurity complexes as $(A^+,A^0) \to
(A^0,A^+)$, $(A^+,A^-) \to (A^0, {\tilde A^0})$, and $({\tilde
A^0}, A^0) \to (A^-,A^+)$. Thus one concludes that the probability
for the first configuration to be included into the aggregate is
given by the product $N_{A^+}N_{A^0}$ while for the second two
configurations - by the product $N_{A^+}^2N_{A^0}^2$. In the
region specified by Eq.~(\ref{NA}), the magnetic field changes the
relation between numbers of the $A^+$- and $A^0$-centers. The
effect of magnetic field on $g_u$ (related to $N_{A^+}$) and $g_l$
(related to $N_{\tilde A^0} = N_{A^0}$) can be estimated from
Eq.~(\ref{A+}) for $r = r_c$.

Since the magnetic field changes the binding energy of an $A^+$ center,
$\delta U_{A^+} = \mu_B g H$, it changes the characteristic distance,
$r_c$. Since $N_{A^+}/N_A \simeq r_c^2 N_A$ this change leads to a
variation in the number of $A^+$: $\delta N_{A^+}{A^+} \simeq r_c^2
N_A$. Having this fact in mind and using Eq.~(\ref{A+}) one obtains:
    \begin{equation}\label{deltaN+} \frac{\delta N_{A^+}}{N_{A^+}} \simeq
    \frac{2\delta U_{A^+} \kappa [r_c^2 + (d_w + d_b)^2/4
      ]^{3/2}}{r_c^2e^2} = \frac{\delta U_{A^+}}{U_{\tilde A^0} - 
    U_{A^+}}\, . \nonumber \end{equation}
 As it was noted above, the
concentration of the aggregates is controlled by the product
$N_{A^+}N_{A^0} = N_{A^+}(N_A - N_{A^+})$. The variation of this
product, in its turn, is
    \begin{equation} \delta N_{A^+}(N_A - 2 N_{A^+})\, . \end{equation}
Therefore one concludes that the effect of
magnetic field depends on the sign of $N_A - 2N_{A^+}$. If this sign is
positive, that is the concentration of $A^+$ centers is relatively
small, the concentration of the aggregates decreases with an application of
magnetic field (decreasing $N_{A^+}$). In its turn, it leads to a
suppression of the polaron effect and to a decrease of the resistance.
If the concentration of $A^+$-centers is large, the magnetic field can
lead to an increase of the resistance. Since the experiment demonstrates
a decrease of resistance as a result of the application of the magnetic
field, we conclude that in our case the density of $A^0$-centers exceeds
the density of $A^+$-centers. This assumption is supported by the fact
that in our experiments the typical distances between acceptors,
$N_A^{-1/2}$, were relatively large, at least $N_A^{-1/2} > (d_w + d_b)$
which is in favor of creating $\tilde A^0$ centers rather than $A^+$centers.

The relative decrease of the density of states of the aggregates,
$\delta \bar P/\bar P \simeq \delta N_{A^+}/ N_{A^+}$ can be estimated
from Eq.~(\ref{deltaN+}) as
    \begin{equation} \label{ef} \frac{\delta
    \bar P}{\bar P} \sim \frac{2 \mu g \langle|H|\rangle}{U_{\tilde A^0}-U_{A^+}}
    \end{equation}
 where $\langle |H|\rangle$ is the time average of the absolute
value of the magnetic field.
This ratio is equal to the relative (with respect to the total number of sites
experiencing the polaron effect) number of the hopping sites where
the polaron cloud is destroyed by the magnetic field. It is
important that this effect is linear in $\langle |H| \rangle$ and
can dominate over the quadratic effect mentioned above.

For further estimates we have to specify the TLS distribution function.
A crude estimate for $\bar P$ in the case of electronic aggregates
is~\cite{ours}
     \begin{equation} \label{P1} {\bar P} = \frac{N\, e^{-
    \lambda N}}{(N\rho^2)N^{1/2}(e^2/\kappa \rho)}\, . \end{equation}
Here
$N$ is the number of pairs of sites forming the aggregate, the
exponential is a statistical factor describing the probability to
construct the bistable aggregate, $\rho$ is the typical distance between
the sites forming the aggregate. Here the first factor in the
denominator describes the typical volume of the (2D) aggregate while the
second - typical scatter of the TLS energy splitting. The factor
$\lambda$ depends on the competition between the Coulomb interactions
within the system and scatter of single-particle energies. Indeed, for
weak Coulomb interactions the system is in its ground state and
occupation of all single-particle states is given. It is the Coulomb
correlations that allow to have metastable configurations with close
total energies. Unfortunately, the number $\lambda$ is not known; it can
depend on realization of the Coulomb glass. However one expects that
large relaxation times are available at not too large $N$ and thus the
exponential is not too small. Assuming $N \approx 5$, $e^{-\lambda N}
\approx 0.1$, $\rho =\xi a \approx 100$ nm (that is of the order of the
typical hopping length) one estimates $\bar P$ as $\sim 10^{23}$
cm$^{-2}$erg$^{-1}$. Substituting this estimate in Eqs.~(\ref{gap}),
(\ref{ef}) and assuming that, $\mu g\langle  H \rangle/(U_{\tilde A^0}
- U_{A^+}) \sim 3 \cdot 10^{-2}$ one obtains
    \begin{equation} \frac{\delta G}{G} \sim
    0.003 \, \ln \frac{t_{\max}}{t_H}\, . \end{equation}
 Here $t_{\max}$ is
the observation time while $t_H$ is given by the inverse sweep rate. This
estimate by order of magnitude agrees with the experimental data. Note
that the present estimate differs from that of~\cite{ours}
where the effect of magnetic field was attributed to partial suppression
of the polaron gap by the magnetic field to an increase of the energy of
the upper state of an aggregate. The suppression is due to influence of
the magnetic field on the $A^+$-centers making some configuration
inaccessible even with account of the correlation energy. However, we
underestimated a possible effect of an increase of the energy of the
lower state involving $A^+$ centers, which can compensate the effect
related to the increase of the energy of the upper level. In this way,
we overestimated the slow-relaxing part of the conductance. On the other
hand, the experimental result for this quantity was also overestimated
because it was not corrected for the temperature fluctuations. We
believe that the corrected experimental results and the suggested
theoretical model are consistent. Namely, the model explains
experimentally observed gradual increase of conductance with time.

 Thus we have discussed two different mechanisms  of slow relaxations
(of magnetoresistance) in  response to time-dependent external
magnetic field, both related to the polaron effect. The first one is
to some extent similar to the effect of the gate
voltage~\cite{Zvi1,Zvi2,Grenet,Zvi3,Zvi4,Zvi5}.  It  is  induced by
magnetic-field-driven shift of the chemical potential. As it was noted
above, this mechanism does not explain the observed gradual increase
of conductance in time.  We expect that this mechanism will be more
pronounced and even dominant in the situation when the induced shift
is larger than the width of the hopping energy band.

The second effect is related to a \textit{direct} dependence of the
``active" sites responsible for  the polaron effect on the magnetic
field. According to our estimates, it is this effect that is
responsible for the experimentally observed slow relaxing
response.

It worth noting that the slow relaxations in response to
variations of the gate voltage are usually not observed in doped
crystalline semiconductors -- the
experiments~\cite{Zvi1,Zvi2,Grenet,Zvi3,Zvi4,Zvi5} were performed
using samples with significant amount of disorder.  We believe
that there two reasons for that: (i) the TLS-induced polaron
effects are much weaker that a direct influence of the gate
voltage on the DOS; (ii) relatively large sweep rates of  the gate
voltage probably lead to pronounced  non-equilibrium  behaviors.
In the present experiments, the direct influence of the magnetic
field sweep is much weaker, and therefore the effects induced by
TLS polarons can be observed.

To conclude, we observed a slow relaxation of magnetoresistance in
response to applied magnetic field in selectively doped p-GaAs-AlGaAs
structures with partially filled upper Hubbard band. We explain this
behavior as related to the properties of the Coulomb glass formed by
charged centers with account of spin correlations, which are sensitive
to external magnetic field.

\section*{Acknowledgments}
We wish to thank Z. Ovadyahu for critical comments and  J.~Bergli for reading the manuscript.


\begin{thebibliography}{99}
\bibitem{end2}  In~\cite{ours} we have erroneously 
estimated of the sweep rate as 40 Oe/s. We apologize for the mistake.
\end{thebibliography}
\end{document}